# An indirect magnetic approach for determining entropy change in first-order magnetocaloric materials

Kun Xu[1,*], Zhe Li[1], Yuan-Lei Zhang[1], Chao Jing[2]

[1]*Department of Physics and Electronic Engineering, Key Laboratory for Advanced Functional and Low Dimensional Materials of Yunnan Higher Education Institute, Qujing Normal University, Qujing 655011, China*

[2]*Department of Physics, Shanghai University, Shanghai 200444, China*

Email: xukun0830@hotmail.com

**Abstract:**

Taking into account the phase fraction during transition for the first-order magnetocaloric materials, an improved isothermal entropy change ($\Delta S_T$) determination has been put forward based on the Clausius-Clapeyron (CC) equation. It was found that the $\Delta S_T$ value evaluated by our method is in excellent agreement with those determined from the Maxwell-relation (MR) for Ni–Mn–Sn Heusler alloys, which usually presents a weak field-induced phase transforming behavior. In comparison with MR, this method could give rise to a favorable result derived from few thermomagnetic measurements. More importantly, we can eliminate the $\Delta S_T$ overestimation derived from MR, which always exists in the cases of Ni–Co–Mn–In and MnAs systems with a prominent field-induced transition. These results confirmed that such a CC–equation-based method is quite practical and superior to the MR-based ones in eliminating the spurious spike and reducing measuring cost.

**Keywords:** Magnetocaloric material; Heusler alloy; Clausius-Clapeyron equation; Maxwell relation

## 1. Introduction

The magnetocaloric effect (MCE), which is the magnetoresponsive ability of refrigerating when a magnetic material is submitted to the external magnetic field variations, has emerged as an attractive option in the generation of energy-efficient cooling technologies. Since the discovery of pseudo-binary $Gd_5Si_2Ge_2$ compound by Pecharsky and Gschneidner in 1997 [1], the MCE became a promising contender to the conventional vapor-compressive refrigerating methods under ambient conditions for its environmental friendliness, higher cooling efficiency and compactness. Since then, giant MCE has also been extensively studied in other intermetallic compounds, such as $MnAs_{1-x}Sb_x$[2], $MnFeP_{1-x}As_x$[3], $MnFeP_{1-x}Si_x$[2], $La(Fe_x,Si_{1-x})_{13}$/ its hydrides [5, 6], and the family of Heusler alloys like Ni-Mn-Ga [7] and Ni-Mn-Z (Z = In, Sn, Sb) [8]. In most cases, the outstanding MCE is due to the involvement of latent heat produced by structural transition in combination with magnetic ordering and changes in the electronic band structure. Thus, entropy change associated with first-order magnetostructural phase transition (FOPT) tends to be much more complicated than the second-order, purely magnetic transition. As is well-known, the characteristic parameters for a magnetocaloric compound are the isothermal entropy change ($\Delta S_T$) and the adiabatic temperature change ($\Delta T_{ad}$) when a magnetic field is applied. To fully exploit the MCE for a given material, it is essential to clarify the evolution mechanism of $\Delta S_T$ or $\Delta T_{ad}$ with temperature (*T*) and magnetic field (*H*). In general, direct or quasi-direct calorimetric methods for determining $\Delta T_{ad}$ are ultimately desirable, but also can be challenging. Hence, the magnetic-measurement-based indirect methods are used to evaluate $\Delta S_T$. According to such a method, the $\Delta S_T$ can be calculated by the Maxwell relation (MR),

$$\Delta S_T = \int_0^H \left(\frac{\partial M}{\partial T}\right)_H dH \approx \sum_i \frac{M_{i+1} - M_i}{T_{i+1} - T_i} \Delta H_i, \qquad (1)$$

using either isothermal magnetization data (MRIT) or isofield magnetization data (MRIF).[9]

However, a large experimental discrepancy led Giguère et al.[10] to claim that, rather than the MR, the Clausius-Clapeyron (CC) equation should apply to calculate the MCE associated with FOPT, since its magnetization curve is not a continuous, derivable function. Soon after, Gschneidner et al.[11] argued that the discrepancy just originates from kinetic effect. Meanwhile, Sun et al.[12] also pointed out that CC equation is just a special case of integrated MR and is inadequate in evaluating the value of $\Delta S_T$ for an incomplete transformation. In addition, recent progress on inverse MCE research of Ni-Mn-Sn alloys has aroused a new round of argument. Zou et al.[13, 14] claimed that the $\Delta S_T$ is seriously overestimated by MRIT in $Ni_{43}Mn_{46}Sn_{11}$ due to the occurrence of a magnetically inhomogeneous martensitic state. On the contrary, Mañosa et al.[15] demonstrated the agreement of obtained $\Delta S_T$ values from both magnetization and calorimetric measurements in the case of $Ni_{50}Mn_{35}Sn_{15}$.

Despite the validity and generality of MR being still under debate as mentioned above, it is undeniable that the direct application of MRIT would lead to unphysical results. Very often, a "spike" is observed in the $\Delta S_T(T)$ curve, especially in the cases of specific Mn-based compounds. For instance, a "colossal" $\Delta S_T$ up to ~ -320

J /kg K for a field change $\Delta H$ = 5 T was obtained in $Mn_{0.997}Fe_{0.003}As$ under the ambient pressure[16]. This value is far above those reported for all known MCE materials and the theoretical prediction. These "colossal" values evoked a large number of investigations, and they were attributed to spurious artifacts due to the incorrect MRIT application. On this basis, practical methods to modify the $\Delta S_T$ obtained from MRIT have been proposed by Liu [17] and Cui [18]. Nevertheless, the former one was claimed to be inadequate to an extreme case of $Mn_{0.99}Cu_{0.01}As$ [19], while the latter one is very complex for practical use. Therefore, this motivates us to explore a proper method to estimate the realistic $\Delta S_T$ value of for FOPT.

Herein, a proper CC equation used here based on isofield magnetization curves (CCIF), regarding phase molar fraction, was introduced to evaluate the $\Delta S_T$. The feasibility of this method was tested on the Heusler alloys of $Ni_{43}Mn_{46}Sn_{11}$, $Ni_{50}Mn_{35}Sn_{15}$, $Ni_{45}Co_5Mn_{36.6}In_{13.4}$ with inverse MCE, as well as MnAs with direct MCE. For comparison, the corresponding $\Delta S_T$ values were also calculated by MR. Our results demonstrated that such a CC-equation-based method that we proposed is superior to the MR-based ones.

## 2. Experimental

Nominal stoichiometry $Ni_{43}Mn_{46}Sn_{11}$ and $Ni_{50}Mn_{35}Sn_{15}$ ingots were prepared by arc melting high-purity Ni, Mn, Sn metals under an argon atmosphere. After melting, the samples were sealed in evacuated quartz tubes and homogenized at 1173 K for 24 h and subsequently ice-water quenched. The crystal structures and phases of samples were verified by X-ray diffraction (Rigaku, Ultima IV) using Cu $K_\alpha$ radiation. The results show that both the $Ni_{43}Mn_{46}Sn_{11}$ and $Ni_{50}Mn_{35}Sn_{15}$ samples are single phase with a $L2_1$ cubic structure (austenite phase) at room temperature. The magnetic characterizations were carried out by a Quantum Design VersaLab magnetometer (VersaLab[TM], 3 Tesla). When dealing with the $Ni_{45}Co_5Mn_{36.6}In_{13.4}$ and MnAs, since both isofield $M(T)$ and $\Delta S_T$ data derived from MRIT are available from previous literatures [2, 20], they were directly used and compared.

The isofield magnetization measurements were carried out as: (i) Cool the sample down from room temperature to $T_-$ (full martensitic state) in the absence of $H$. (ii) With applying a constant $H$, the measurement is made on increasing $T$ from $T_-$ to $T_+$ (full austenite state) with a scan rate of 1.0 K/min, and the *isofield $M(T)$ heating curve* is obtained. (iii) Without removing $H$, the measurement is taken on the decreasing temperature from $T_+$ to $T_-$, and the *isofield $M(T)$ cooling curve* is obtained.

The isothermal magnetization measurements were performed as: (i) *Isothermal $M_T(H)$ curve on the heating branch*: heat the sample at zero fields from $T_-$ to $T_i$ without overshooting the target temperature, and then record the magnetization $M(H)|_{T=T_i}$ by scanning fields from 0 to 3 T. Then, the sample is cooled down to $T_-$ at zero field, and heat to the next higher temperature $T_{i+1}$. (ii) *Isothermal $M(H)$ curve on the cooling branch*: cool the sample down from $T_+$ to $T_i$ without undershooting the target temperature, and collect the isotherm $M(H)|_{T=T_i}$ by scanning $H$ from 0 to 3 T. Then, the sample is heated up to $T_+$ at zero fields, and cool down to the next lower temperature $T_{i+1}$. Before each subsequent measurement, the sample was carefully recovered to the initial state, since the first-order system could remain partially transformed in higher magnetization after removing the saturate field if the cooling martensitic transformation is not crossed [19, 21].

## 3. Results and discussion

Practically, the isothermal magnetization curve may not reveal the complete FOPT due to the insufficient magnetic field span provided by the apparatus. For example, the field necessary for a complete transition for $Ni_{50.4}Mn_{34}Sn_{15.6}$ is estimated to be as large as ~15.2 T) [22], which is far beyond the capacity of the commercial instrument. Instead, for the common first-order materials with giant MCE, the transition temperature usually locates between 100 K and 350 K. Hence, the temperature-induced transforming process can be captured by the thermomagnetic curves. This suggests that the CCIF method, on the basis of $M(T)$ rather than $M(H)$ curves, is superior in this aspect. To illustrate the CCIF method use, an intensively studied ternary $Ni_{43}Mn_{46}Sn_{11}$ alloy is selected as a representative. Fig. 1 shows its $M(T)$ heating curve from $T_- = 150$ K to $T_+ = 240$ K in the presence of a static field $H = 1$ T. The sample undergoes a reverse martensitic transformation (MT) from a weak-magnetic martensite (*M*) to a ferromagnetic austenite (*A*) phase. To our knowledge, the finite width of the transforming region is due to the inhomogeneous nature of the sample [23], and the transforming process can be viewed as continuous. On this basis, the change of phase fraction ($\Delta f$) has been taken into account to investigate the reported $\Delta T_{ad}$ and hysteresis behaviors associated with FOPT by numerous authors [22-26]. In the isofield condition, the fraction of *A* phase as a function of temperature can be denoted as $f(T)$. The kinetic arrest effect [27], which only occurs at very high fields, is ignored here. The $f(T)$ rises continuously from 0 % to 100 % for the full reverse MT. In this line of reasoning, the correlation between $M(T)$ and $f(T)$ during FOPT process should be established. Indeed, the $f(T)$ can be characterized using magnetization as a proxy by the following expression:

$$f(T) = \frac{M(T) - M_M(T)}{M_A(T) - M_M(T)}, \qquad (2)$$

where $M_M(T)$, $M_A(T)$ represent the magnetizations of full $M$ and $A$ phases, respectively. $M_M(T)$ (or $M_A(T)$) can be roughly estimated by extrapolating the linear part of $M(T)$ curves towards higher (or lower) temperatures, shown as a blue (or red) dot line. Following Eq. (2), Fig. 1 illustrates the resultant $f(T)$ under $H=1$ T by a green solid line.

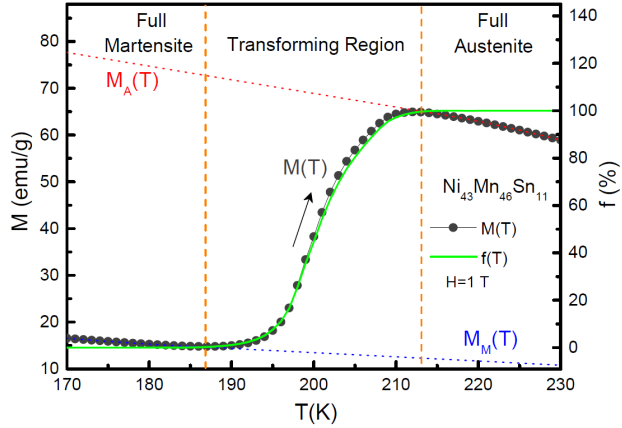

**Fig. 1.** Temperature dependence of magnetization under a static magnetic field of 1 T for Ni$_{43}$Mn$_{46}$Sn$_{11}$ sample (full circles). The red and blue dot lines represent $M_A(T)$ and $M_M(T)$ obtained by extrapolating the linear part of magnetization. The corresponding austenite molar fraction $f(T)|_{H=1T}$ is represented using a green solid line.

Considering the fact that both $T$ and $H$ are thermodynamically equivalent driving forces for the FOPT [22], the impact of $H$ on the $f(T)$ is necessary to be quantitatively analyzed in this paragraph. As Fig. 1 shows, a broad transition (~ 20 K) in $f(T)$ curve is observed, indicating that only a partial phase transition will take place by applying a moderate $H$, hence the potential entropy change cannot be triggered completely [23]. To obtain the exact amount of transformed phase fraction $\Delta f$ triggered by a field variation $\Delta H$ from the initial field $H_i$ to the final field $H_f$, $M(T)$ curves on heating under $H = H_i$, $H_f$ were measured and plotted in the inset of Fig. 2. The $M(T)$ data are transformed to $f(T)$ (see Fig. 2) through the same approach mentioned above. Consequently, the $\Delta f$ can be computed by the expression $\Delta f(T) = f(T)|_{H=H_f} - f(T)|_{H=H_i}$. Moreover, the CC equation,

$$\Delta S_{tr} = -\Delta M \frac{dH}{dT_t}, \qquad (3)$$

can describe the ideal FOPT directly. Additionally, the only drawback is that it is justified only in the case of complete magnetic-field-induced transformation [21]). Here, $\Delta M$ is the jump of magnetization across the transition, and $\Delta S_{tr}$ is the total entropy change for the fully driven transition (*i.e.* the fact $\Delta S_T \approx \Delta S_{tr}$ [28, 29] when $\Delta f$ reaches 100 %). By taking $\Delta f$ into account, which reflects the degree of phase transformation completion percent, we can determine the $\Delta S_T$ for an incomplete magnetic-field induced transformation by using CC equation. Here, the CCIF method to evaluate $\Delta S_T$ has been well established by the following equation,

$$\Delta S_T \approx \Delta f \Delta S_{tr} = -\Delta f \Delta M \frac{dH}{dT_t}. \qquad (4)$$

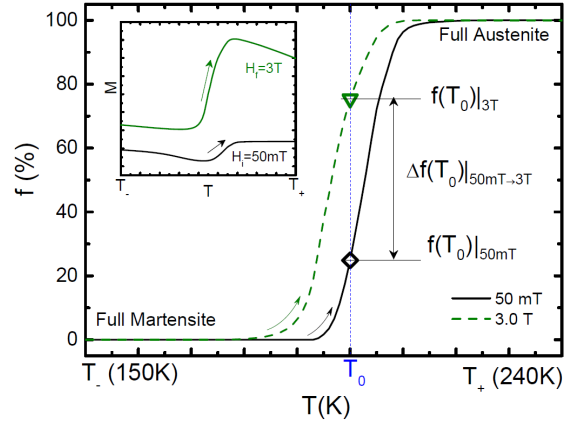

**Fig. 2.** The schematic illustration of calculated $f(T)$ curves of $Ni_{43}Mn_{46}Sn_{11}$ under fields of $H_i$ and $H_f$ in the heating run. The hollow diamond and inverse triangle represent the $f(T_0)|_{H=50mT}$ and $f(T_0)|_{H=3T}$. Inset: The $M(T)$ under fields of 50 mT and 3 T.

To test the validity of this CCIF approach for Ni-Mn-Sn system, the $\Delta S_T$ of $Ni_{43}Mn_{46}Sn_{11}$ and $Ni_{50}Mn_{35}Sn_{15}$ alloys in both cooling and heating regimes will be derived from CCIF, MRIT, and MRIF approaches, respectively. Sec. II describes the magnetization measurements procedures under isothermal and isofield conditions. For $Ni_{43}Mn_{46}Sn_{11}$, the $T_-$ and $T_+$ were chosen to be 150 K and 240 K. Fig. 3 presents the $M(T)$ isofield curves at different magnetic fields up to 3 T. The thermal hysteresis of ~10 K indicates the first-order nature of this transition. Its $\Delta M$ magnitude is almost kept constant (~50.7 emu/g on heating and ~51.0 emu/g on cooling) with increasing $H$, as the inset of Fig. 3 shows. The $T_t$, corresponding to the middle point in magnetization change, increases almost linearly with $H$ at a rate of ~ -2.03 and ~ -2.28 K/T under the heating and cooling modes (see Fig. 4 (a)). Thus, the $\Delta S_{tr}$ of $Ni_{43}Mn_{46}Sn_{11}$ is obtained to be ~24.97 and ~22.37 J/kg K under the two modes using CC equation. The $\Delta f$ for the field changes from $H_i$ (50 mT) to $H_f$ (1 T, 2 T, 3 T) can be computed by subtracting two corresponding $f(T)$ curves. Figs. 5 (a) and (b) demonstrate the $\Delta S_T$ curves estimated from CCIF method by multiplied the constant value of $\Delta S_{tr}$. At the same time, $\Delta S_T$ values from the MRIT (Figs. 5 (c) and (d)) and the MRIF (Figs. 5 (e) and (f)) have been also derived for comparison. During the $\Delta S_T$ determination by means of the MRIF, more $M(T)$ data were collected and added into computation to increase the reliability and accuracy of results. As seen from these results, due to the small value of $dT_t/dH$ and a weak field-induced MT, the spurious spike does not appear in the determination from MRIT. The comparison highlights the fact that the peak values calculated by CCIF agree with those from MR. However, for the cooling protocol, the $\Delta S_T$ peaks derived from temperature-scan data (CCIF and MRIF) are centered at a temperature ~3 K lower than the one from isotherms (MRIT). This small shift can be probably attributed to the thermal inertia of calorimeter which may affect isofield measurements [26]. In addition, it is worth noting that, although the resulted $\Delta S_T$ from these three approaches reach quite similar values, all of $\Delta S_T$ in the cooling run are actually overestimated since the hysteresis effect was neglected. In the cooling path of MT, an increase of $H$ will convert it into reverse MT, thus a great part of $H$ will be consumed to overcome the hysteresis subsequently. This explains why in the cooling run a considerable $\Delta S_T$ from MRIF, whereas a much weaker $\Delta T_{ad}$, from direct measurement were observed in the $Ni_{50}Mn_{36}Co_1Sn_{13}$ system [9]. Such complex path-dependent phenomenon in the first-order systems still requires further quantitative study to obtain a more realistic $\Delta S_T$ value for practical use.

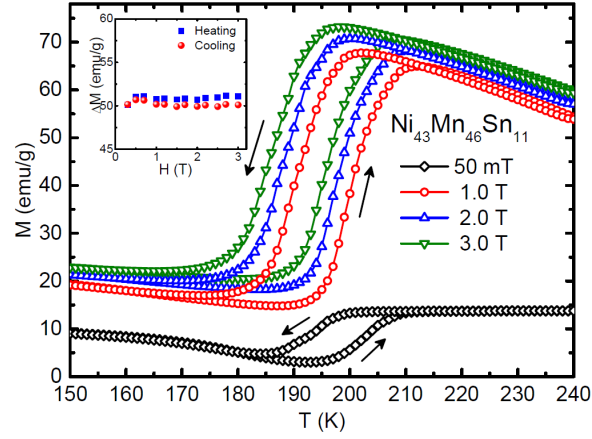

**Fig. 3.** $M(T)$ isofield curves of $Ni_{43}Mn_{46}Sn_{11}$ upon heating and cooling in magnetic fields of 50 mT, 1 T, 2 T and 3 T. Inset: Field dependence of $\Delta M$.

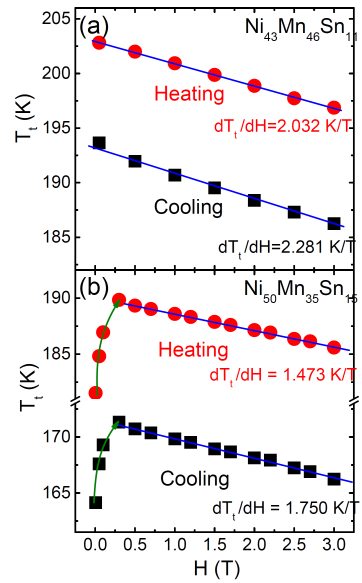

**Fig. 4.** $T_t$ versus $H$ phase diagram deduced from isofield curves for (a) $Ni_{43}Mn_{46}Sn_{11}$ and (b) $Ni_{50}Mn_{35}Sn_{15}$. The solid lines are guides to eye.

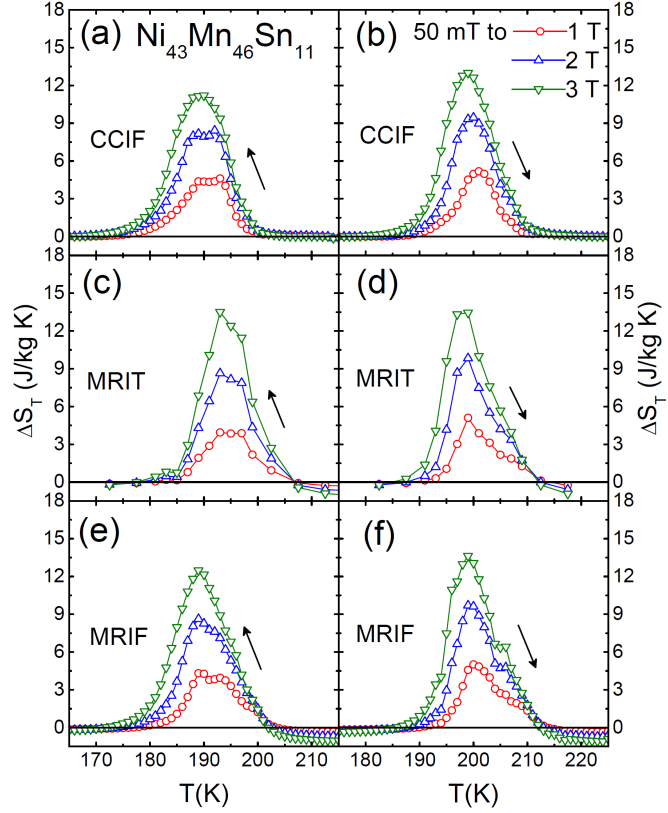

**Fig. 5.** The calculated $\Delta S_T(T)$ of $Ni_{43}Mn_{46}Sn_{11}$ (a)-(b) by CCIF; (c)-(d) by MRIT; (e)-(f) by MRIF. The left panel shows those upon cooling, and the right panel shows those for heating.

Next, the same approach was implemented on the $Ni_{50}Mn_{35}Sn_{15}$ alloy, where $T_-$ and $T_+$ were chosen to be 120 K and 230 K, respectively. Compared to $Ni_{43}Mn_{46}Sn_{11}$, a larger thermal hysteresis (~20 K) and a smaller saturate $\Delta M$ (~28 emu/g for heating and ~31 emu/g for cooling) are obtained (*not shown here*). The $\Delta M$ value is strongly restrained, because the $M$ phase is in its ferromagnetic state at transition, which is detrimental to gain a large $\Delta S_T$. Then, the parameter $dT_t/dH$ should be derived to fulfill the calculation. However, as Fig. 4 (b) shows, the linear relation fails in the fields below 0.3 T, which has been also observed by Shamberger *et al*. [22] and Barandiaran *et al*. [30] This means that the CCIF is incapable to evaluate $\Delta S_T$ of Ni$_{50}$Mn$_{35}$Sn$_{15}$ over the whole field span. For this reason, the $H_i$ and $H_f$ have to be chosen in the linear region, where the values of $dT_t/dH$ are derived to be ~ -1.47 and ~ -1.75 K/T under the two modes. The corresponding $\Delta S_{tr}$ (~19.05 and ~17.71 J/kg K, respectively) are smaller than the previous case. Following the approach as used in Ni$_{43}$Mn$_{46}$Sn$_{11}$, $\Delta S_T$ for field changes from $H_i$ (0.5 T) to $H_f$ (1 T, 2 T, 3 T) can be calculated, and Fig. 6 shows the comparisons among CCIF, MRIT and MRIF for Ni$_{50}$Mn$_{35}$Sn$_{15}$. The peak values of $\Delta S_T$ are almost coincident, and a more pronounced temperature discrepancy (~6 K) is observed between isofields and isotherms in the cooling run. The larger temperature shift should be attributed to larger thermal hysteresis (~20 K) and broader measuring temperature span (~110 K, larger than 90 K in Ni$_{43}$Mn$_{46}$Sn$_{11}$).

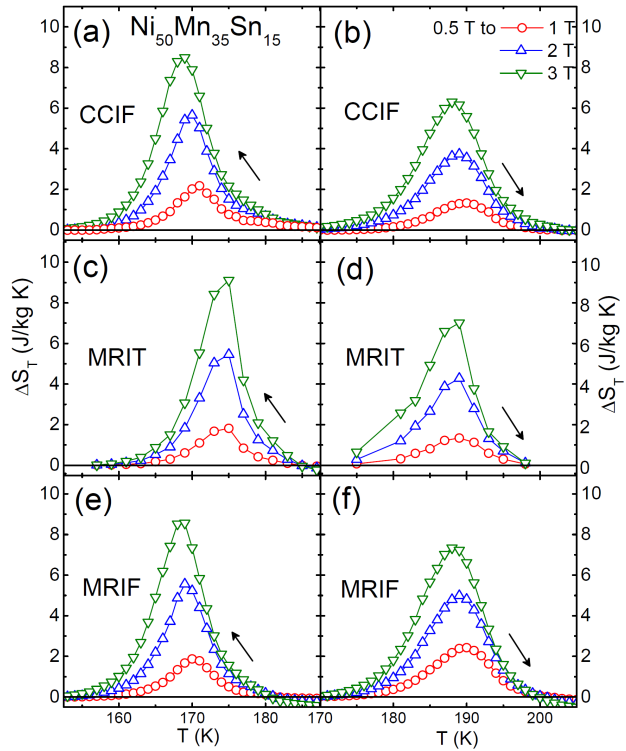

**Fig. 6.** The calculated $\Delta S_T(T)$ of Ni$_{50}$Mn$_{35}$Sn$_{15}$ (a)-(b) by CCIF; (c)-(d) by MRIT; (e)-(f) by MRIF. The left panel shows those upon cooling, and the right panel shows those for heating.

Previous works concluded that a large $\Delta M$ and a high sensitive $T_t$ as a function of the $H$ contribute to enhance the inverse MCE, and they are strongly interlinked. According to Eq. (4), the former one is proportional to the $\Delta S_{tr}$, and also plays a key role on the MCE. Meanwhile, a larger $\Delta S_{tr}$ requires a smaller $dT_t/dH$, which often causes partial phase transition (small $\Delta f$) and weakens the $\Delta S_T$ product in turn. Overall, the transition span must be minimized and $dT_t/dH$ should be compromised to a moderate value to gain a giant $\Delta S_T$ over a wide temperature range [23, 28]. In these two cases of Ni-Mn-Sn, relative small $dT_t/dH$ together with broad transition span led them to have a relatively large $\Delta S_{tr}$ but only with a small $\Delta f$ (less than 50% for 3 T), and further a peak-shape $\Delta S_T$ curve.

Besides the ternary Ni–Mn–Z alloys, suitable elements (Co, Fe, etc.) were always doped to tune the $T_t$ or alter the magnetic coupling between Mn–Mn atoms to exhibit excellent performances in a variety of applications[20, 31-33]. For instance, Chen *et al.* [20] reported the further advancements of MCE and magnetoresistance in the $Ni_{50}Mn_{36.6}In_{13.4}$ system with Ni partially substituted by both Co and Fe. A spurious spike was observed in the $\Delta S_T$ curve from MRIT due to a prominent field-induced transforming behavior in this system. During their effective refrigerant capacity evaluation, the high plateau instead of the spike value was taken as the effective $\Delta S_T$ value. However, this way of choosing the effective $\Delta S_T$ value by cutting the spike is somewhat arbitrary and may be infeasible in some extreme cases, i.e. when the plateau is completely covered by the spurious spike. In order to estimate its realistic value of $\Delta S_T$ where the spike appears and confirm the reliability of the CCIF, the $\Delta S_T$ value of quaternary alloy Ni$_{45}$Co$_5$Mn$_{36.6}$In$_{13.4}$ for a $\Delta H = 5$ T (using $\Delta M = 97.49$ emu/g and $dT_t/dH = 5.0$ K/T) is recalculated and compared with the data given in Ref. 20 by MRIT, as Fig. 7 shows. Interestingly, it is found that the result from CCIF is in excellent agreement with MRIT once again, and the $\Delta S_T(T)$ curve from CCIF form a plateau with the spike disappearing. Moreover, it is obvious to find that the $\Delta S_T(T)$ curve from CCIF is over a wider temperature range. The difference originates from the fact that the total entropy change has been put into a very narrow temperature interval for the results from MRIT[19].

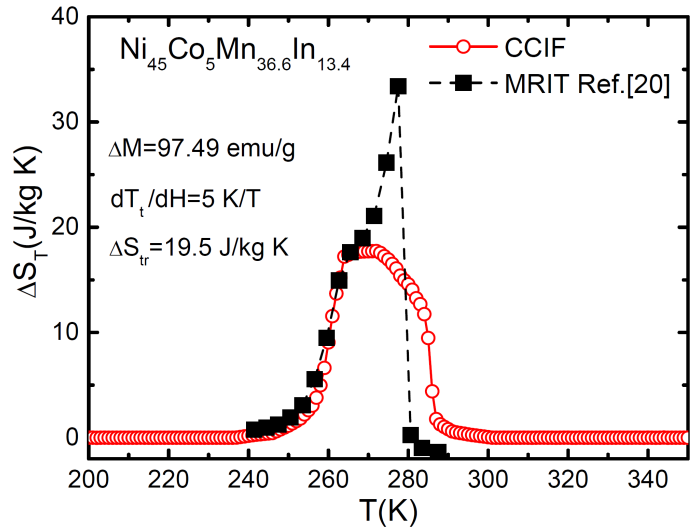

**Fig. 7.** The $\Delta S_T(T)$ for a field change $\Delta H = 5$ T of Ni$_{45}$Co$_5$Mn$_{36.6}$In$_{13.4}$ by CCIF (open circles) and MRIT (full rectangles). The original $M(T)$ data and $\Delta S_T$ by MRIT are referred to Ref. 20.

These representatives discussed above indicate that the CCIF is a quite simple and effective way to determine $\Delta S_T$ in the particularly interesting class of Ni-Mn-Z with inverse MCE, and the shortcoming of MRIT inadequacy in the phase-separated state, is overcome. Besides, the CCIF need fewer $M(T)$ curves than MRIF to provide the same accuracy. It requires only a few $M(T)$ curves as long as there is enough to obtain the value of $dT_t/dH$.

To these ends, we try to apply this method to the direct first-order magnetocaloric materials. A representative direct magnetocaloric compound with numerous disputations, MnAs, was selected to test the generality of this method. For MnAs and MnAs$_{1-x}$Sb$_x$ systems, the transition no longer involves $M$ or $A$ phases, but a low-temperature ferromagnetic phase with the hexagonal MnP-type structure and a high-temperature paramagnetic phase with the orthorhombic NiAs-type structure. Let $f(T)$ be the molar fraction of high temperature phase, and $f(T)$ can be expressed as follows:

$$f(T) = \frac{M(T) - M_{LT}(T)}{M_{HT}(T) - M_{LT}(T)}, \tag{5}$$

where $M_{LT}(T)$ and $M_{HT}(T)$ represent the magnetizations of low- and high-temperature phases, respectively.

The $M(T)$ isofield data in the heating run are extracted from Ref. 2. A striking value of saturate $\Delta M$ (~95.5 emu/g) and an extremely narrow transition temperature span (less than 5 K) are observed. For the MnAs system, $T_t$ is more sensitive to $H$ than in the Ni-Mn-Sn system, and shifts to higher $T$ with a rate of ~3.4 K/T. Thereby, a "moderate" $\Delta S_{tr}$, ~28.09 J/kg K, is achieved. In this case, the remarkable $dT_t/dH$ value and narrow transition span ensure a much larger $\Delta f$ than the one with small $dT_t/dH$ and wide transition span (in Ni-Mn-Sn system) for the same $\Delta H$. Specifically, the phase transformation completion percentage ($\Delta f$) ~78% and ~98%, can be achieved for just a small $\Delta H$ of 0–1 T and for a moderate $\Delta H$ of 0–2 T. It warrants a nearly complete transition over a wide range in temperature, and forms a plateau in both $\Delta f$ and $\Delta S_T(T)$ curves. As a consequence, a giant refrigerant capacity could be achieved. The two sets of the $\Delta S_T(T)$ data, evaluated by the CCIF and the MRIT (the data of Fig. 2 of Ref. 2), are in a nearly quantitative agreement except in the phase-mixed region, as Fig. 8 displays. The unreasonable spike is diminished, and the problem of overestimating $\Delta S_T$ in the vicinity of $T_t$ is also resolved.

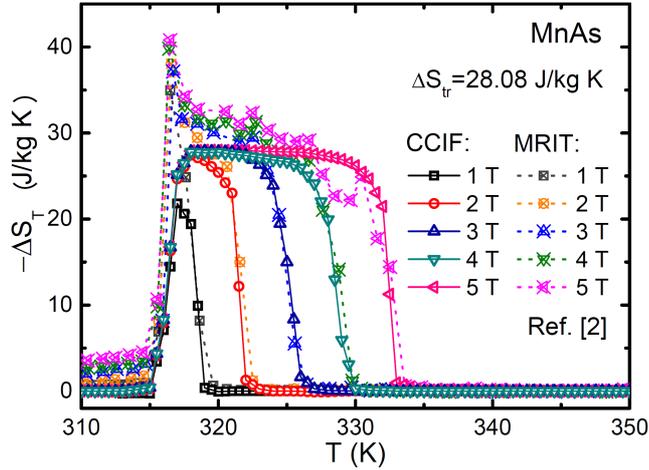

**Fig. 8.** The $\Delta S_T(T)$ of MnAs by both CCIF (full lines) and MRIT (dot lines). The original data and $\Delta S_T$ by MRIT are referred to Ref. 2.

## 4. Conclusions

Despite the FOMT and MCE behaviors of Ni-Mn-Z alloy having been investigated for more than one decade, the most commonly used approach in evaluating $\Delta S_T$, the MRIT, still has the problem that a spurious "spike" appears in the $\Delta S_T(T)$ curve. Based on the CC equation, a relative simple method dealing with $M(T)$ data, in the aspect of phase fraction change, was developed to determine the MCE features in the present study. It has been shown that the results of the three methods (CCIF, MRIT, MRIF) converged to the same value in Ni-Mn-Z system. Moreover, avoiding the appearance of the unwilling "spike", and requiring few magnetization curves make the CCIF a powerful tool for a reliable qualitative $\Delta S_T$ estimation. On the other hand, an application of this method to the direct MCE MnAs system also achieved excellent results. In the future, this approach can be generalized to other direct/ inverse MCE materials with FOPT, and encourages the appearance of novel MCE material by controlling the key parameters $\Delta M$, $dT_t/dH$ as well as transition width.

**Acknowledgements**

This work has been carried out with financial support of the National Natural Science Foundation of China (No. 11404186, 11364035, 51371111), the Youth Program of Fundamental Application of Yunnan Province (No. 2012FD051), and the Bid Project of Qujing Normal University (No. 2011ZB001). We thank LetPub (www.letpub.com) for its linguistic assistance during the preparation of this manuscript.

**References**

[1] V.K. Pecharsky, K.A. Gschneidner, Jr., Physical Review Letters 78 (1997) 4494.

[2] H. Wada, Y. Tanabe, Applied Physics Letters 79 (2001) 3302.

[3] O. Tegus, E. Brück, K.H.J. Buschow, F.R. de Boer, Nature 415 (2002) 150.

[4] X.F. Miao, L. Caron, P. Roy, N.H. Dung, L. Zhang, W.A. Kockelmann, R.A. de Groot, N.H. van Dijk, E. Brück, Physical Review B 89 (2014) 174429.

[5] F.X. Hu, B.G. Shen, J.R. Sun, Z.H. Cheng, G.H. Rao, X.X. Zhang, Applied Physics Letters 78 (2001) 3675.

[6] A. Fujita, S. Fujieda, Y. Hasegawa, K. Fukamichi, Physical Review B 67 (2003) 104416.

[7] M. Pasquale, C.P. Sasso, L.H. Lewis, L. Giudici, T. Lograsso, D. Schlagel, Physical Review B 72 (2005) 094435.

[8] T. Krenke, E. Duman, M. Acet, E.F. Wassermann, X. Moya, Ll. Mañosa, A. Planes, Nature Materials 4 (2005) 450.

[9] V.V. Khovaylo, S.P. Skokov, O. Gutfleisch, H. Miki, T. Takagi, T. Kanomata, V.V. Koledov, V.G. Shavrov, G. Wang, E. Palacios, J. Bartolomé, R. Brurriel, Physical Review B 81 (2010) 214406.

[10] A. Giguère, M. Foldeaki, B. Ravi Gopal, R. Chahine, T.K. Bose, A. Frydman, J.A. Barclay, Physical Review Letters 83 (1999) 2262.

[11] K.A. Gschneidner, Jr., V.K. Pecharsky, E. Brück, H.G.M. Duijn, E.M. Levin, Physical Review Letters 85 (2000) 4190.

[12] J.R. Sun, F.X. Hu, and B.G. Shen, Physical Review Letters 85 (2000) 4191.

[13] J.D. Zou, B.G. Shen, B. Gao, J. Shen, J.R. Sun, Advanced Materials 21 (2009) 693.

[14] J.D. Zou, B.G. Shen, B. Gao, J. Shen, J.R. Sun, Advanced Materials 21 (2009) 3727.

[15] Ll. Mañosa, A. Planes, X. Moya, Advanced Materials 21 (2009) 3725.

[16] A. de Campos, D.L. Rocco, A.M.G. Carvalho, L. Caron, A.A. Coelho, S. Gama,

L.M. da Silva, F.C.G. Gandra, A.O.D. Santos, L.P. Cardoso, P.J. von Rank, N.A. de Oliveira, Nature Materials 5 (2006) 802.

[17] G.J. Liu, J.R. Sun, J. Shen, B. Gao, H.W. Zhang, F.X. Hu, B.G. Shen, Applied Physics Letters 90 (2007) 032507.

[18] W.B. Cui, W.Liu, Z.D. Zhang, Applied Physics Letters 96 (2010) 222509.

[19] L. Caron, Z.Q. Ou, T.T. Nguyen, D.T. Cam Thanh, O. Tegus, E. Brück, Journal of Magnetism and Magnetic Materials 321 (2009) 3559.

[20] L. Chen, F.X. Hu, J. Wang, L.F. Bao, J.R. Sun, B.G. Shen, J.H. Yin, L.Q. Pan, Applied Physics Letters 101 (2012) 012401.

[21] G. Porcari, S. Fabbrici, C. Pernechele, F. Albertini, M. Buzzi, A. Paoluzi, J. Kamarad, Z. Arnold, M. Solzi, Physical Review B 85 (2012) 024414.

[22] P.J. Shamberger, F.S. Ohuchi, Physical Review B 79 (2009) 144407.

[23] J. Liu, T. Gottschall, K.P. Skokov, J.D. Moore, O. Gutfleisch, Nature Materials 11 (2012) 620.

[24] E. Stern-Taulats, P.O. Castillo-Villa, Ll. Mañosa, C. Frontera, S. Pramanick, S. Majumdar, A. Planes, Journal of Applied Physics 115 (2014) 173907.

[25] V. Basso, C.P. Sasso, K.P. Skokov, O. Gutfleisch, V.V. Khovaylo, Physical Review B 85 (2012) 014430.

[26] B. Emre, S. Yüce, E. Stern-Taulats, A. Planes, S. Fabbrici, F. Albertini, Ll. Mañosa, Journal of Applied Physics 113 (2013) 213905.

[27] W. Ito, K. Ito, R. Y. Umetsu, R. Kainuma, K. Koyama, K. Watanabe, A. Fujita, K. Oikawa, K. Ishida, T. Kanomata, Applied Physics Letters 92 (2008) 021908.

[28] V. Recarte, J.I. Pérez-Landazábal, S. Kustov, E. Cesari, Journal of Applied Physics 107 (2010) 053501.

[29] V.K. Pecharsky, K.A. Gshneidner, Jr., A.O. Pecharsky, A.M. Tishin, Physical Review B 64 (2001) 144406.

[30] J.M. Barandiaran, V.A. Chernenko, E. Cesari, D. Salas, P. Lazpita, J. Gutierrez, I. Orue, Applied Physics Letters 102 (2013) 071904.

[31] L. Feng, L. Ma, E.K. Liu, G.H. Wu, W.H. Wang, W.X. Zhang, Applied Physics Letters 100 (2012) 152401.

[32] M. Khan, I. Dubenko, S. Stadler, J. Jung, S.S. Stoyko, A. Mar, A. Quetz, T. Samanta, N. Ali, K.H. Chow, Applied Physics Letters 102 (2013) 112402.

[33] X.G. Zhao, M. Tong, C.W. Shih, B. Li, W.C. Chang, W. Liu, Z.D. Zhang, Journal of Applied Physics 113 (2013) 17A913.